\documentclass[aps,prl,twocolumn,floats]{revtex4}
\usepackage{graphicx}
\usepackage{graphicx}
\usepackage{times}
\usepackage{natbib}

\begin{document}
\title{Chemical and Mechanical Signaling in Epithelial Spreading}
\author{L. M. Pismen$^1$ and S. Y. Shvartsman$^2$}
\affiliation{$^1$ Department of Chemical Engineering and Minerva Center for Nonlinear Physics
of Complex Systems, Technion -- Israel Institute of Technology, Haifa, Israel \\
$^2$ Department of Chemical Engineering and Lewis--Sigler Institute for Integrative
Genomics,\\ Princeton University, Princeton NJ, USA}

\begin{abstract}
We propose a minimal mathematical model to explain long-range coordination of
dynamics of multiple cells in epithelial spreading, which may be induced, under different conditions, by a chemical signal, or mechanical stress, or both. The model is based on chemo-mechanical interactions including a chemical
effect of stress and concentration-dependent traction. The results,
showing kinase concentration distribution and cell displacement, allow us to reproduce two activation waves on different time scales observed in the experiment, and distinguish between distinct dynamical patterns observed under conditions of injury or unconstraining.
\end{abstract}
\maketitle

Regulated movement and deformation of epithelial layers, two-dimensional sheets of cells that tightly adhere to each other, is a key process in adult and developing tissues. During embryogenesis, spreading and folding of epithelia plays a central role in gastrulation, an event that initiates the formation of three-dimensional structures of tissues and organs. In adult tissues, epithelial migration is important for wound healing. Mechanisms of collective cell migrations are a subject of intense research  \cite{Friedl09}. It is generally believed that epithelial spreading involves polarization of the cell cohort propagating from ``leading'' cells to the ``followers''. The process is mediated by the intracell actin machinery  and proteins of adherens junctions, and regulated by complicated chemical pathways, which still remain largely unexplored. 

Early models of epidermal wound healing \cite{Murray91}, adopted also in a number of later publications, were based on reaction-diffusion equations describing cell motion and proliferation in response to a diffusive chemical signal secreted by the wound. This approach turned out to be inadequate for the description of spreading epithelia, which has to include the effect of both chemical and mechanical factors on collective cell motion. Recent \emph{in vitro} studies gave evidence of the importance of chemo-mechanical interactions, whereby directional cell migration is maintained by a positive feedback loop between cell deformation and intracellular signal transduction. Matsubayashi \emph{et al} \cite{Matsu04} proposed that collective cell migration during wound healing is caused by an activation (phosphorylation) wave of Mitogen Activated Protein Kinase (MAPK). The experiments of Nicolic \emph{et al}  \cite{woundheal} using canine kidney cells followed up on this by tracing both spreading dynamics and distribution of activated kinase levels under different boundary conditions. 

\begin{figure}[b]
\includegraphics[width=8.5cm]{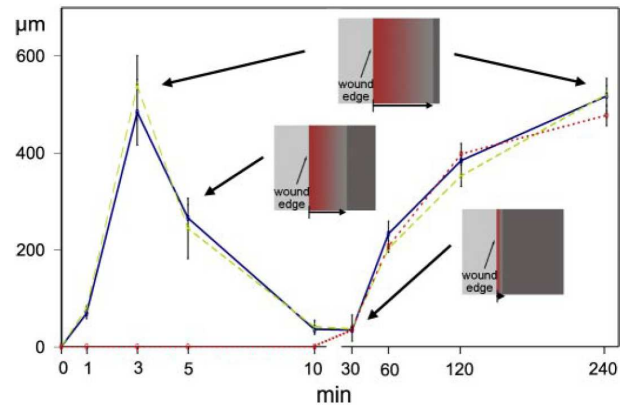} 
\caption{\label{f1} 
Distance from the wound edge that a MAPK wave reaches at different
time points after injury (scratching, solid line or membrane peel-off, dashed line) and unconstraining without injury (dotted line). In the latter case,
only one wave is present. Insets: schematic representation of the MAPK waves reaching different distances at different times  \cite{woundheal}.}
\end{figure}

Wounding generates both free space for spreading cells, and an active species emitted as a result of injury that initiates the spreading process. Spreading also occurs, however, albeit at a somewhat slower speed, in the absence of injury, being initiated by a purely mechanical action following unconstraining of the epithelial layer. In the case of spreading initiated by injury, two waves of MAPK activation were observed (Fig.~\ref{f1}). The first fast wave is excited by the signal emitted at the wounded edge; consequently, it rapidly decays, and is followed by a slow wave gradually spreading into the interior of the epithelial layer. In the absence of injury, only the slow wave is present; its form and spreading speed are similar to those in injured epithelium, indicating that the same mechanical signal is likely to be operating in both cases. The experiment confirmed as well the crucial role of MAPK, proved by suppression of spreading in the presence of a MAPK inhibitor \cite{woundheal}.

Very recently, Trepat  \emph{et al} \cite{cellmig09} reported direct measurements of cell traction and stress in a spreading epithelial layer (in the absence of injury), which proved that traction is not restricted to a proximity of the leading edge, as has been commonly believed, but remains substantial deep inside the layer. This implies an effective mechanism of propagation of active traction within the tissue. In this Letter, we propose that directional collective cell migration is maintained by a positive feedback loop between cell deformation and intracellular signal transduction. Thus, mechanical and/or chemical signals at the wound edge can coordinate shape changes and movements of multiple cells deep within the epithelium.

Our aim is to find out and test a minimal phenomenological model compatible with available observations. While chemical pathways and chemo-mechanical interactions are still unknown (and are certainly far more complex than the rough model to be described), there are few well-pronounced experimental features that need to be explained: (a) faster advance in the presence of injury, especially at the leading edge \cite{woundheal}; (b) activation of MAPK in the absence of injury, and similarity of the emerging MAPK wave with the second wave in the presence of injury \cite{woundheal}; (c) strong internal traction, large internal stress \cite{cellmig09} and weak decay of the speed of advance with depth \cite{woundheal} in uninjured epithelia.

Our main task is to discern a minimal set of chemo-mechanical interactions that would allow one to explain the above features. We will concentrate therefore on estimating forces and displacement velocities averaged in the direction parallel to the leading edge, and abstract at this stage from a strong random component of the cell motion \cite{silberzan,cellmig09} causing stress and velocity fluctuations, as well as transverse instabilities which naturally arise when  positive feedback loops are operative \cite{Cross,book}.  

Since the motion is strongly overdamped, the appropriate equation governing the displacement velocities $v_i$ should be based on Aristotelean rather than Newtonian dynamics, and has the form similar to the Darcy law:
\begin{equation}  
\label{darcy}  
v_i = K f_i,  \qquad f_i = \partial_j \sigma_{ij}.
\end{equation}
Here $f_i $ is the force vector, which can be defined in a usual way through the stress tensor $\sigma_{ij}$; $K$ is the mobility coefficient, or inverse friction factor, and $\partial_j=\partial/\partial x_j$ is the spatial derivative; summation over repeated indices is presumed. Although the mobility should be generally anisotropic in a polarizable medium \cite{Joanny}, this is irrelevant in the current context of the study of averaged motion where the only non-vanishing components are those in the direction $x$ normal to the edge of the tissue, $f=f_x$ and $\sigma=\sigma_{xx}$.

The forces acting in the epithelial tissue are at least of three kinds. First, is an external  ``wetting'' force $f^{(w)}$ applied at the leading edge, responsible for a tendency of an unconstrained layer to spread to available space. The other passive force is due to the tissue elasticity. The simplest model is the Hooke's law $f^{(e)}=E \,\partial_x u$ where $u$ is the displacement along the $x$ axis and $E$ is the elastic constant. Finally, there is an active traction force $f^{(a)}$ arising within the cells.

MAPK activation, caused by either chemical or mechanical signaling, plays a crucial role in cell spreading, as indicated by suppression of spreading by MAPK inhibition \cite{woundheal}. Although the mechanism of MAPK action remains so far unclear, it is reasonable to assume that it enables active traction, so that the traction force can be assumed to be proportional to the level of MAPK activation $m$, $f^{(a)} = F m$. The experiment suggests that MAPK should be activated both chemically and mechanically, but mechanical activation is likely to be delayed, to account for the observed retardation of the MAPK wave in the absence of injury (Fig.~\ref{f1}). We postulate therefore that MAPK is activated directly by the signal emitted by the injury at the leading edge, but its mechanical activation occurs through an intermediate chemical species, which is produced by straining stress. 

Since we strive to keep the model as simple as possible, we will assume linear kinetic relationships whenever it is possible. One should, however, be cautious to avoid runaway in the model containing a positive feedback loop. Such a loop is created here by the stress arising due to the motion driven by MAPK-activated traction, which, in turn, activates more MAPK. The runaway instability is prevented by assuming the dependence of the intermediate species production rate on stress to be saturable. Thus, we assume the following kinetic model governing the concentrations of the signal, $s$, and intermediate species, $c$, and the level of MAPK activation, $m$:   
\begin{eqnarray}  
\label{signal}  
\dot s &=& D \partial^2_x s -  s, \\
\label{int}  
\dot c &=&  \frac{a \sigma H(\sigma)}{1+b \sigma} - q c, \\
\label{mapk}  
\dot m &=&  s + c - k m, 
\end{eqnarray}
where $H(\sigma)$ is the Heaviside step function. The only diffusive species in this scheme is the signal, with the diffusivity $D$; the dot denotes the derivative with respect to the time variable, which is made dimensionless by using the inverse signal decay constant $1/k_s$ as the time scale; other decay constants are actually ratios, $q =k_i/k_s$ and $k=k_m/k_s$. The MAPK activation constants in Eq.~(\ref{mapk}) are eliminated by adopting suitable relative concentration scales. 

Another problem to be addressed in model building is that of spreading of stress within the tissue. Recent experiments \cite{cellstress08} suggest that stress does not propagate inside cells or tissues instantaneously or with the speed of sound, but in a delayed fashion, showing both appreciable delay and attenuation with distance, which may be compatible with the poroelastic model \cite{stressMaha08}. 
Lacking reliable data, we stopped at the elastic model (\ref{darcy}) enhanced by advective propagation of traction against the direction of motion \cite{cellmig09}. Accordingly, the continuum version of the mechanical model (\ref{darcy}), collecting the elastic force $f^{(e)}$ and traction $f^{(a)}$, reads
\begin{equation}  
\label{force}  
\dot u = -\chi (m - \alpha\partial_x m) + \gamma\, \partial_x^2 u, 
\end{equation}
where $\chi= KF, \; \gamma=KE$, and $\alpha$ is the fraction of the cell traction transmitted backwards into the tissue, and the traction is assumed to be negatively directed. The suitably scaled stress is computed by integrating the r.h.s.\ of  Eq.~(\ref{force}):
\begin{equation}  
\label{stress}  
\sigma = f^{(w)} +\chi \left(\int_0^x m \, dx - \alpha m \right) 
- \gamma( \partial_x u -1). 
\end{equation}
The spatially discretized one-dimensional discrete version of Eqs.~(\ref{force}), (\ref{stress}) to be solved below is
\begin{eqnarray}    
\dot u_n &=& - \chi_n [m_n - \alpha( m_{n+1} -m_n)] \nonumber \\
 &+& \gamma ( u_{n+1} -2u_n + u_{n-1}),  \label{move} \\
\sigma_n &=& f^{(w)} +\chi \left(\sum_{j=1}^n m_j - \alpha m_{n+1} \right) 
\nonumber \\ &-& \gamma\,(u_{n+1} -u_n -1).
\label{movs} \end{eqnarray}
where $u_n, \; n=0, \ldots ,N$ is specified as the position of the front of $n$th cell, and the effective mobility $\chi_n$ is allowed to be position-dependent. 

The first stage, present in a wounded tissue only, is chemical activation by a pulse emitted at the leading edge and diffusing into the tissue. The diffusion is very fast on the time scale of cell motion, reaching the depth of dozens of cells in a matter of minutes (see Fig.~\ref{f1}), and therefore can be treated without taking any reshaping of the tissue or its cellular structure into consideration. In this approximation, the solution of Eq.~(\ref{signal}) in the half-plane $x>0$ for a semi-infinite tissue with the instantaneous pulse at the leading edge $s(x,0) =\delta(x)$ as the initial condition and the no-flux boundary condition at the leading edge $x=0$ is
\begin{equation}  
\label{sign}  
s(x,t) = \frac{1}{\sqrt{\pi D t}} \exp \left( - t -\frac{x^2}{4Dt} \right).
\end{equation}
The solution is normalized by the total injected signal, so that $\int_0^\infty s(x,t) \, dx =e^{- t}$.
The solution for the level of MAPK activation, neglecting at this stage the mechanical contribution, can be also expressed analytically: 
\begin{eqnarray}  
& & m(x,t) =\frac{ e^{-k t}} {2 \sqrt{D(1-k)}} \times \hfill \nonumber \\ 
& & \left\{e^{-x \sqrt{(1-k)/D} } \left[ \text{erf}\left( \sqrt{(1-k)t} -  \frac{x}{\sqrt{2 Dt}}  \right)+1 \right] +
 \right.  \nonumber \\   & & \left.
   e^{x \sqrt{(1-k)/D} } \left[ \text{erf}\left( \sqrt{(1-k)t} +  \frac{x}{\sqrt{2 Dt}}  \right) -1 \right]  \right\}, 
\label{mapkc}    \end{eqnarray} 
where erf(...) is the error function. The analytical expressions (\ref{sign}), (\ref{mapkc}) are applicable in the Lagrangian frame with the coordinate $x$ referring to their initial positions. Both the signal level and the level of MAPK activation pass through a maximum with time at any location.  Another possible assumption of a constant production rate above a certain threshold restricts MAPK to a finite band with a width first growing and then shrinking with time, and dynamics of its concentration within this band is qualitatively similar to the above. 

\begin{figure} 
\begin{tabular}{c}
\includegraphics[width=7.2cm]{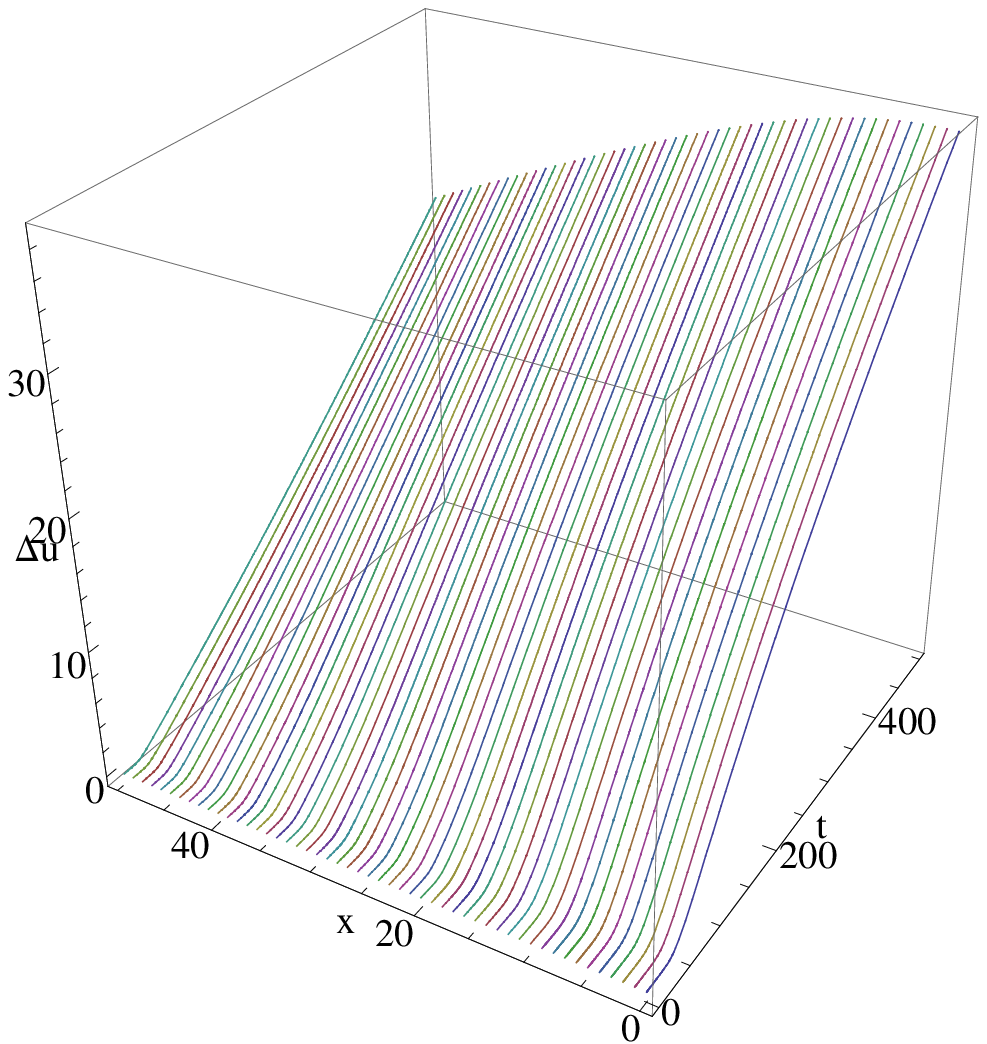} \\
\includegraphics[width=7.2cm]{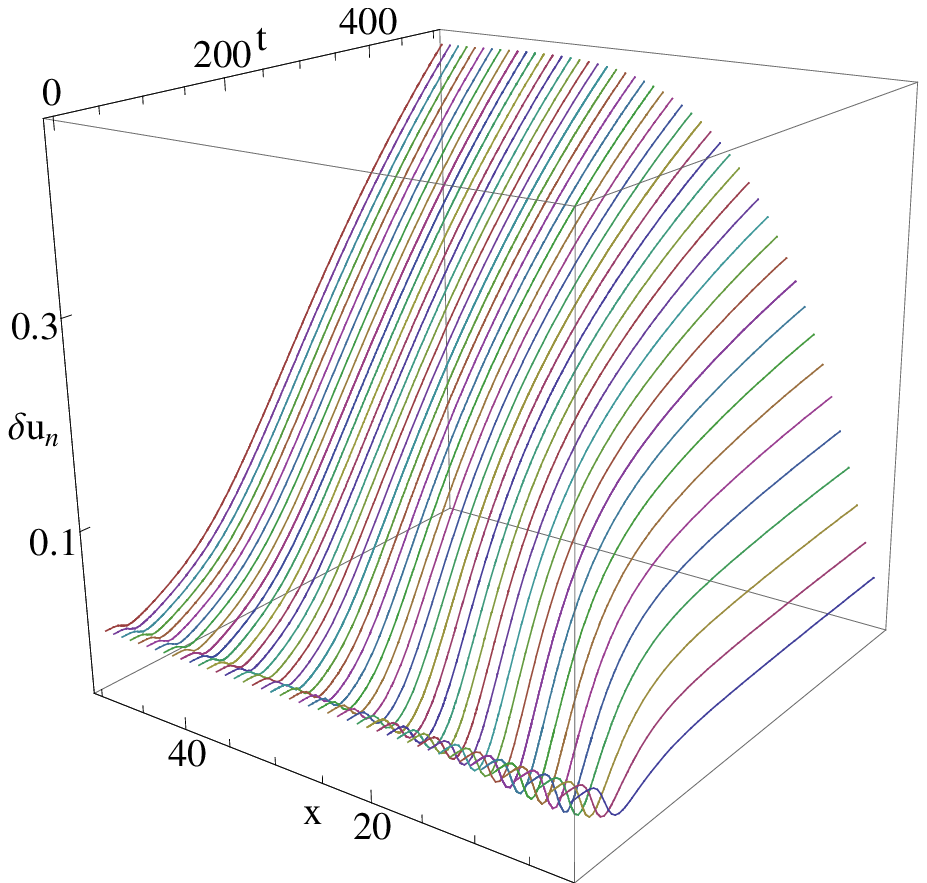} \\
\includegraphics[width=7.2cm]{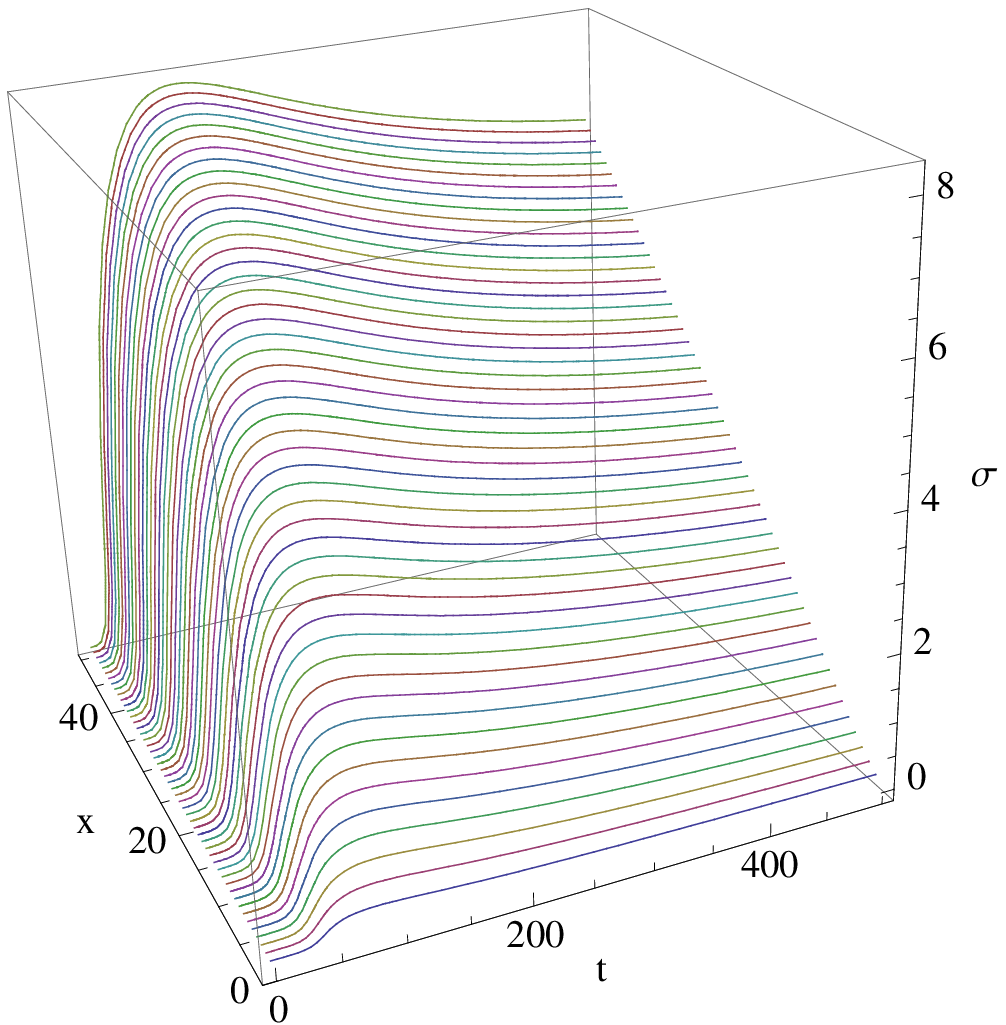} 
\end{tabular} \caption{\label{fr1} 
The cell displacement $\Delta u_n$ (above), extension $\delta_n=u_n-u_{n-1}-1$ (middle) and stress (below) in the absence of injury in a 50-cell chain. The distances are measured in the units of cell size $ \lambda $. The parameters: $D=400\, \lambda^2, \;  \gamma=5\, \lambda^{-1}, \;  k_0=0.1, \; k_1=2,\;  q=0.05, \; \chi= 0.5 \, \lambda \;  a=0.01,\; b=1,\;  \alpha=0.2, f^{(w)}=10^{-3}$.}
\end{figure}

As follows from from experiment \cite{woundheal} and integrating Eq.~(\ref{move}) using $m(x,t)$ given by  Eq.~(\ref{mapkc}) with the decay constants ratio $k \approx 0.1$ that fits the dynamics in Fig.~\ref{f1}, the advance at the purely chemical stage (ending with the decay of both the signal species and signal-activated MAPK) is limited to a few cell lengths. This feable advantage cannot account for the observed substantial difference between overall long-time cell drifts with and without injury, which is most pronounced close to the scratched edge. To emulate this difference, we had to modify the model  by assuming that traction is enhanced in the regions reached by the signal, e.g. due to accumulation of an additional very slowly decaying promoter, so that $\chi=\chi_0(1 + \chi_1 S)$ with 
\begin{equation}  
 S(x,t)=\int_0^t s(x,t) \,dt , \quad \lim_{x \to \infty} S(x,t)= \frac{ 1} {2 \sqrt{D}} e^{-x /\sqrt{D} } .
\label{mapks}    \end{equation} 

\begin{figure}
\includegraphics[width=8cm]{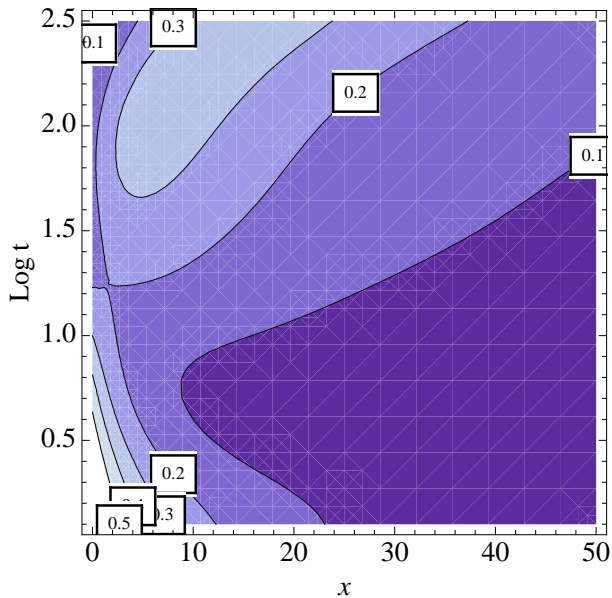} 
\caption{\label{fcm} 
The map of the level of MAPK activation in spreading following injury$ (\chi_1=5$, other parameters as in Fig.~\ref{fr1}.}
\end{figure}

\begin{figure}
\includegraphics[width=8cm]{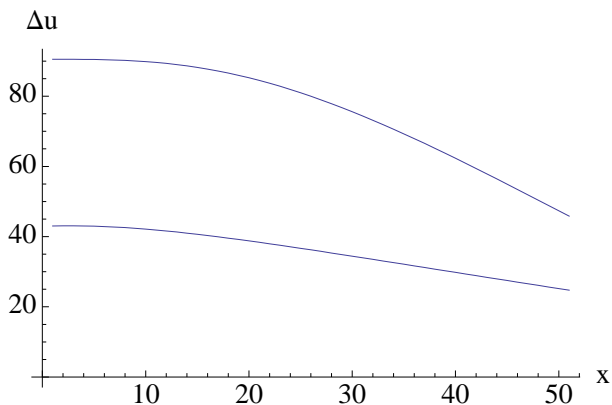}
\caption{\label{fdu} 
Coomparison of cell displacement at the end of computation runs with (upper curve) and without (lower curve) injury.  }
\end{figure}

Another modification of the basic model we found to be necessary is acceleration of the MAPK decay under the influence of stress; otherwise excessive drift is observed in the back where large stress accumulates in our computations, as well as in the experiment \cite{cellmig09}. Thus, we set in Eq.~(\ref{mapk}) $k=k_0(1 + k_1 \sigma)$. Under this assumption, the model contains actually a feedforward loop connecting the stress and the level of MAPK activation: the former activates the latter through the intermediate species and suppresses it directly. 

We integrated Eq.~(\ref{move}) together with Eq.~(\ref{movs}) and the discretized Eqs.~(\ref{signal})--(\ref{mapk}) for a chain of 50 cells with the continuation boundary conditions at the back end imitating a semi-infinite layer. The computations were carried out with the help of the \emph{Mathematica} ODE solver. In the absence of injury, the solution of Eqs.~(\ref{int})--(\ref{force}) with zero boundary conditions at the leading edge and zero initial conditions is trivial, and a weak ``wetting'' force $f^{(w)}$ applied at the leading edge is necessary to initiate motion, which is consequently accelerated by the positive feedback loop. An example of computation starting from an unperturbed tissue is given in Fig.~\ref{fr1}. The stress is accumulated in the back, in accordance with the data of Trepat  \emph{et al} \cite{cellmig09}. The extension is also maximal at the back, which may explain that it is there that most cell divisions occur \cite{silberzan}, even though divisions, a relatively rare event, are not incorporated in this model. The advance approaches a quasistationary speed in the model semi-infinite tissue, though, of course, it would stop in a finite tissue with suppressed cell divisions.

With an injury, the stress-dependent activation of MAPK follows the signal-induced stage. In accordance with the experiment \cite{woundheal}, the the level of MAPK activation, after reaching a maximum, decreases before going up again, as seen in Fig.~\ref{fcm}. The difference between the drift of front and back cells is here, in accordance with the experiment, far more pronounced than in the absence of injury. The displacements at the end of computation runs with identical parameters are compared in Fig.~\ref{fdu}.

 
We anticipate that the above minimal model built under conditions of uncertainty will encourage further quantitative studies on chemo-mechanical interactions in spreading tissues, and will, in its turn, be adjusted and modified as more data  become available.

\acknowledgments{This collaboration is supported by the Human Frontier Science Program (Grant RGP0052/2009-C)}.


\begin{thebibliography}{17}
\bibitem{Friedl09} P.~Friedl and  D.~Gilmour, Nat. Rev. Mol. Cell Biol., \textbf{10}, 445 (2009)
\bibitem{Murray91} R.~T.~Tranquillo and  J.~D.~Murray, J.~Theor.~Biol,  \textbf{158}, 135 (1992).
\bibitem{Matsu04} Y.~Matsubayashi, M.~Ebisuya, S.~Honjoh, and E.Nishida, Curr.\ Biol.,  \textbf{14} 731 (2004).
\bibitem{woundheal}  D.~L.~Nikolic, A.~N.~Boettiger, D.~Bar-Sagi, J.~D.~Carbeck and S.~Y.~Shvartsman, Am.~J.~Physiol.~Cell Physiol., \textbf{291} C68 (2006).
\bibitem{cellmig09} X.~Trepat, M.~R.~Wasserman, T.~E.~Angelini, E.~Millet, D.~A.~Weitz, J.~P.~Butler,  and J.~J.~Fredberg, Nat. Phys., \textbf{5} 426 (2009).
\bibitem{silberzan} M.~Poujade,  E.~Grasland-Mongrain, A.~Hertzog, J.~Jouanneau, P.~Chavrier, B.~Ladoux, A.~Buguin, and P.~Silberzan, PNAS, \textbf{104} 15988 (2007).
\bibitem{Cross} M.\ C.\ Cross,  and P.\ Hohenberg, Rev.\ Mod.\ Phys.\ \textbf{65}, 851 (1993).
\bibitem{book} L.~M.~Pismen, Patterns and Interfaces in Dissipative Dynamics, Springer Series in Synergetics (2006).
\bibitem{Joanny} J.~F.~Joanny, F.~Juelicher, K.~Kruse, and J.~Prost, New J.\ Phys., \textbf{9} 422 (2007).
\bibitem{cellstress08} M.J.~Rosenbluth, A.~Crow, J.~W.~Shaevitz, and D.~A.~Fletcher,  Biophys.J., \textbf{95} 6052 (2008).
\bibitem{stressMaha08} T.~J.~Mitchison, G.~T.~Charras, and L.~Mahadevan, Semin.\ Cell Dev.\ Biol., \textbf{19} 215 (2008).

\end{thebibliography}
\end{document}